%% file: main.tex
\documentclass[aps, prd, twocolumn, floats, floatfix, superscriptaddress, nofootinbib]{revtex4}

\usepackage[utf8]{inputenc}
\usepackage{amsfonts,amsmath,amssymb,amsthm,mathtools}
\usepackage[dvipsnames,x11names]{xcolor}
\usepackage[sort&compress]{natbib}

\usepackage{graphicx}
\usepackage{booktabs}
\usepackage{array}

\usepackage{latexsym}
\usepackage[citecolor=MidnightBlue,linkcolor=MidnightBlue,colorlinks=true]{hyperref}
\usepackage[title]{appendix}
\usepackage{aas_macros}
\usepackage{natbib}

\usepackage{siunitx}

\usepackage[normalem]{ulem}

\usepackage{afterpage}

\def\colorstar{black}
\def\colorH{black}
\def\colorms{black}
\def\colorext{black}
\def\colorstd{black}

\def\MS{{\color{\colorstar}{M_S}}}
\def\OS{{\color{\colorstar}{\Omega_S}}}
\def\IS{{\color{\colorstar}{I_S}}}
\def\QS{{\color{\colorstar}{Q_S}}}
\def\JS{{\color{\colorstar}{J_S}}}
\def\lS{{\color{\colorstar}{\lambda_S}}}
\def\chiS{{\color{\colorstar}{\chi_S}}}

\def\MbH{{\color{\colorH}{M_0^H}}}
\def\RbH{{\color{\colorH}{R_0^H}}}

\def\QH{{\color{\colorH}{Q^H}}}

\def\ISH{{\color{\colorH}{I_S^H}}}
\def\QSH{{\color{\colorH}{Q_S^H}}}

\def\dMH{{\color{\colorH}{\delta M{}^H}}}
\def\lH{{\color{\colorH}{\lambda_S^H}}}
\def\MSH{{\color{\colorH}{M_S^H}}}
\def\OKH{{\color{\colorH}{\Omega_K^H}}}

\def\Mbms{{\color{\colorms}{M_0^\star}}}
\def\Rbms{{\color{\colorms}{R_0^\star}}}
\def\Mms{{\color{\colorms}{M_S^\star}}}
\def\lms{{\color{\colorms}{\lambda_S^\star}}}
\def\Oms{{\color{\colorms}{\Omega_S^\star}}}
\def\Ims{{\color{\colorms}{I_S^\star}}}
\def\Qms{{\color{\colorms}{Q_S^\star}}}
\def\Kms{{\color{\colorms}{K^\star}}}
\def\gammams{{\color{\colorms}{\gamma^\star}}}
\def\Pcms{{\color{\colorms}{P_c^\star}}}
\def\chims{{\color{\colorms}{\chi_S^\star}}}

\def\pms{{\color{\colorms}{p^\star}}}

\def\Bms{{\color{\colorms}{B^\star}}}

\def\ext{{\color{\colorext}{\text{ext}}}}

\def\miI{{\color{\colorext}{I_S^{\ext}}}}
\def\miQ{{\color{\colorext}{Q_S^{\ext}}}}
\def\miM{{\color{\colorext}{M_0^{\ext}}}}

\def\std{{\color{\colorstd}{\text{std}}}}

\def\suI{{\color{\colorstd}{I_S^{\std}}}}
\def\suQ{{\color{\colorstd}{Q_S^{\std}}}}
\def\suM{{\color{\colorstd}{M_0^{\std}}}}

\def\errors{\mathcal{E}}
\def\mierror{{\color{\colorext}{\errors^{\ext}}}}
\def\mierrorvector{{\color{\colorext}{\vec{\errors}^{\ext}}}}

\def\mierrorK{{\color{\colorext}{\errors_{K}^{\ext}}}}
\def\mierrorgamma{{\color{\colorext}{\errors_{\gamma}^{\ext}}}}
\def\mierrorPc{{\color{\colorext}{\errors_{P_c}^{\ext}}}}

\def\mierrorI{{\color{\colorext}{\errors_{I_S}^{\ext}}}}
\def\mierrorQ{{\color{\colorext}{\errors_{Q_S}^{\ext}}}}
\def\mierrorM{{\color{\colorext}{\errors_{M_0}^{\ext}}}}

\def\suerrorK{{\color{\colorstd}{\errors_{K}^{\std}}}}
\def\suerrorgamma{{\color{\colorstd}{\errors_{\gamma}^{\std}}}}
\def\suerrorPc{{\color{\colorstd}{\errors_{P_c}^{\std}}}}

\def\suerrorI{{\color{\colorstd}{\errors_{I_S}^{\std}}}}
\def\suerrorM{{\color{\colorstd}{\errors_{M_0}^{\std}}}}
\def\suerrorQ{{\color{\colorstd}{\errors_{Q_S}^{\std}}}}

\def\NOTATIONS{{\color{\colorstar}{X_S}}}
\def\NOTATIONms{{\color{\colorms}{X^\star}}}
\def\NOTATIONH{{\color{\colorH}{X^H}}}
\def\NOTATIONmsEOS{{\color{\colorms}{\mathrm{EoS}^\star}}}
\def\NOTATIONext{{\color{\colorext}{X^{\text{ext}}}}}
\def\NOTATIONerrorext{{\color{\colorext}{\errors_{X}^{\ext}}}}
\def\NOTATIONstd{{\color{\colorstd}{X^{\text{std}}}}}

\def\mipunto{{\color{\colorext}{p^{\ext}}}}
\def\miradio{{\color{\colorext}{\radio^{\ext}}}}

\def\Mb{M_0}
\def\Rb{R_0}

\def\deltaM{\delta M}

\def\radio{\mathcal{R}}

\def\radiolms{{\color{\colorms}{\mathcal{R}^{\star}_{\lambda_S}}}}
\def\radioMms{{\color{\colorms}{\mathcal{R}^{\star}_{M_S}}}}
\def\radiomiI{{\color{\colorext}{\mathcal{R}^{\ext}_{I_S}}}}
\def\radiomiQ{{\color{\colorext}{\mathcal{R}^{\ext}_{Q_S}}}}
\def\radiomiM{{\color{\colorext}{\mathcal{R}^{\ext}_{M_0}}}}

\def\Curve{\mathcal{C}}

\def\Clove{{\color{\colorms}{\mathcal{C}_{\lambda_S}^{\star}}}}
\def\CMS{{\color{\colorms}{\mathcal{C}_{M_S}^{\star}}}}
\def\CMb{{\color{\colorms}{\mathcal{C}_{M_0}^{\star}}}}
\def\CI{{\color{\colorms}{\mathcal{C}_{I_S}^{\star}}}}
\def\CQ{{\color{\colorms}{\mathcal{C}_{Q_S}^{\star}}}}
\def\CMbR{{\color{\colorext}{\mathcal{C}_{M_0}^{\ext}}}}
\def\CIR{{\color{\colorext}{\mathcal{C}_{I_S}^{\ext}}}}
\def\CQR{{\color{\colorext}{\mathcal{C}_{Q_S}^{\ext}}}}
\def\CMbW{{\color{\colorstd}{\mathcal{C}_{M_0}^{\std}}}}
\def\CIW{{\color{\colorstd}{\mathcal{C}_{I_S}^{\std}}}}
\def\CQW{{\color{\colorstd}{\mathcal{C}_{Q_S}^{\std}}}}

\def\Slove{{\color{\colorms}{\mathcal{S}_{\lambda_S}^{\star}}}}
\def\SMS{{\color{\colorms}{\mathcal{S}_{M_S}^{\star}}}}
\def\SMb{{\color{\colorms}{\mathcal{S}_{M_0}^{\star}}}}
\def\SI{{\color{\colorms}{\mathcal{S}_{I_S}^{\star}}}}
\def\SQ{{\color{\colorms}{\mathcal{S}_{Q_S}^{\star}}}}
\def\SMbR{{\color{\colorext}{\mathcal{S}_{M_0}^{\ext}}}}
\def\SIR{{\color{\colorext}{\mathcal{S}_{I_S}^{\ext}}}}
\def\SQR{{\color{\colorext}{\mathcal{S}_{Q_S}^{\ext}}}}
\def\SMbW{{\color{\colorstd}{\mathcal{S}_{M_0}^{\std}}}}
\def\SIW{{\color{\colorstd}{\mathcal{S}_{I_S}^{\std}}}}
\def\SQW{{\color{\colorstd}{\mathcal{S}_{Q_S}^{\std}}}}

\def\pol{\mathrm{Pol}}
\def\indi{y}

\newcommand{\todo}[1]{\textcolor{Magenta}{#1}}

\newcolumntype{C}{>{\centering\arraybackslash}m{6em}}

\newcounter{mnotecount}%
\newcommand{\mnote}[1]%
{\protect{\stepcounter{mnotecount}}$^{\mbox{\footnotesize $\bullet$\themnotecount}}$ 
\marginpar{%
\raggedright\tiny\textit{
$\!\!\!\!\!\!\,\bullet$\themnotecount: #1} }}

\makeatletter
\newcommand*{\textvcenter}{\@ifstar{\@tempswatrue\text@vcenter}{\@tempswafalse\text@vcenter}}
\newcommand*{\text@vcenter}[1]{\mbox{$\m@th\vcenter{\setbox\z@=\hbox{#1}\if@tempswa\dp\z@\z@\fi\box\z@}$}}

\newcommand*{\showmathaxis}{%
   \setbox\z@=\hbox{$a$}%
   \@tempdima\fontdimen22\textfont2\relax
   \@tempdimb\@tempdima
   \advance\@tempdima.2\p@
   \advance\@tempdimb-.2\p@
   \leavevmode\rlap{\vrule height\@tempdima depth-\@tempdimb width10cm}%
}
\makeatother

\begin{document}

\input{v_def}

\bibliography{super_fluid}
\bibliographystyle{ieeetr}

\end{document}

%% file: v_def.tex
\title{\boldmath 
$I$-Love-$Q$, and $\deltaM$ too: The role of the mass in universal relations of compact stars}

\author{Eneko Aranguren}
\email{eneko.aranguren@ehu.eus}
\affiliation{Department of Physics, University of the Basque Country UPV/EHU, P.O.~Box 644, 48080 Bilbao, Basque Country, Spain}
\author{Jos\'e A.~Font}
\email{j.antonio.font@uv.es}
\affiliation{Departamento de Astronom\'ia y Astrof\'isica, Universitat de Val\`encia, Dr. Moliner 50, 46100, Burjassot (Valencia), Spain}
\affiliation{Observatori Astron\`omic, Universitat de Val\`encia, Catedr\'atico Jos\'e Beltr\'an 2, 46980, Paterna, Spain}
\author{Nicolas Sanchis-Gual}
\email{nicolas.sanchis@uv.es}
\affiliation{Departamento de Astronom\'ia y Astrof\'isica, Universitat de Val\`encia, Dr. Moliner 50, 46100, Burjassot (Valencia), Spain}
 \author{Ra\"ul Vera}
\email{raul.vera@ehu.eus}
\affiliation{Department of Physics, University of the Basque Country UPV/EHU, P.O.~Box 644, 48080 Bilbao, Basque Country, Spain}

\begin{abstract}
In the study of rotating neutron stars the $I$-Love-$Q$ relations refer to the existence of various approximate, equation of state-independent relations involving
the moment of inertia, the Love number and the quadrupole moment. 
These relations are relevant for observational astrophysics, since 
they allow (in theory) the inference of any two quantities within the $I$-Love-$Q$ triad out of the third one alone.
However, the quantities involved in the relations are, in fact, normalized by a parameter $\Mb$ that arises in the usual perturbative analytical approach as the mass of
the background configuration. Since $\Mb$ is not the mass of the rotating star $\MS$, it is not 
an observational quantity, which may affect the application of the relations to actual observations.
This situation is usually ignored in most studies by taking $\Mb$ to be the mass of the star, an approximation that can, in some cases, be inconsistent. In this paper we extract the value of $\Mb$ using an {\it extended} version of the universal relations that involve a fourth parameter, $\deltaM$, proportional to the difference $\MS-\Mb$. We analyze to which degree this extended set of relations yields a more precise inference of compact star properties and equation of state parameters.
\end{abstract}

\maketitle

\section{Introduction}

In the field of relativistic astrophysics, semi-analytical solutions of the Einstein Field Equations for isolated and rotating compact stars are usually computed using perturbation theory to second order. This approach makes  possible to obtain some properties of the star, such as the moment of inertia $I$ and the quadrupole moment $Q$, for a given equation of state (EoS), along with the central pressure and angular velocity. The perturbative approach also yields the mass of the static configuration $\Mb$, derived from solving the Tolman-Oppenheimer-Volkoff (TOV) equation, and the mass of the rotating star $\MS$, to second order. The former is a purely mathematical result, with no actual observational implication, 
whereas the latter provides the actual mass of the rotating star (within the accuracy inherent to the second order approximation). The difference between \todo{$\MS$} and $\Mb$ is
proportional to a quantity, usually denoted as $\deltaM$, computed at second order.
The proportionality factor is the square of the
stellar angular velocity, $\OS$ (see e.g. \cite{reina-sanchis-vera-font2017}).
Static configurations embedded in a quadrupolar tidal field can also be explored semi-analytically within a
perturbative approach to calculate the Love number \cite{Hinderer:2008,Flanagan2008,Hinderer2009}, which aims at describing the effect of a companion star in a binary system.

The discovery of approximate (EoS-insensitive) universal relations among $I$, Love, and $Q$~\cite{Yagi:2013, Yagi:2013awa, Yagi:2014},
to which we will refer in the following simply as ``universal relations", has been widely claimed to be useful in observational astrophysics. 
As stated in \cite{Yagi:2013awa}, the measurement of a single member of the $I$-Love-$Q$ trio would automatically allow to compute the remaining two parameters solely using the relations, without any further measurement. Hence, the $I$-Love-$Q$ universal relations are a most useful tool to derive neutron star properties and constrain the EoS of matter at supranuclear densities.

These universal relations, however, do not involve the bare quantities $I$ and $Q$ but some dimensionless counterparts
$\overline{I}$ and $\overline{Q}$, which are normalized with the
(not directly observable) TOV mass $\Mb$. Consequently,
to extract the values of the bare quantities from the universal 
relations, one needs, in principle, to somehow fix $\Mb$.
The approach usually considered in the literature assumes
that $\Mb$ is close to the actual stellar mass $\MS$, which is a
direct observable.
In fact, the standard approach takes $\Mb=\MS$,
which is explicitly acknowledged as a ``small caveat" in \cite{Yagi:2013awa} and justified for neutron stars rotating with dimensionless spin parameters $\chiS\lesssim0.1$.
Those spins, however, are far from the values observed in
millisecond pulsars, which can attain values as large as $\chiS\sim  0.4$ (PSR J1748-2446ad~\cite{Hessels:2006}). As we discuss in Section~\ref{sec:conclusions} below, recent works have argued if the absence of millisecond pulsars in binary neutron star (BNS) systems might actually be the result of an observational bias~\cite{Papenfort:2021,Dudi:2022,Papenfort:2022,Rosswog:2023}. This motivates a reassessment of the implications of the $\Mb=\MS$ approximation in the adequacy of the universal relations.
Such a reevaluation was in fact already discussed in our previous investigation~\cite{reina-sanchis-vera-font2017} where additional universal relations  were reported. Those involve also $\deltaM$, turning the $I$-Love-$Q$ trio into a quartet. It was also shown in~\cite{reina-sanchis-vera-font2017} that this extended set allows to unambiguously extract $\Mb$ with no approximation (and thus the bare $I$ and $Q$ quantities). It should be noted that rotation-dependent ``universal" relations, which would apply to rapidly rotating stars as well, have also been reported~\cite{Doneva:2013rha,Pappas:2013naa,Chakrabarti:2013tca}. However, the extended $I$-Love-$Q$-$\deltaM$ universal relations discussed in~\cite{reina-sanchis-vera-font2017} and here sidestep the use of such rotation-dependent relations by enabling the inference of $\Mb$ in the first place.

The aim of the present paper is to test to which extent the use of the standard $I$-Love-$Q$ relations to extract the non-normalized corresponding values can be enhanced by using the extended set of relations (including $\deltaM$). We do it in two ways. First, we check numerically the relative errors introduced by
the standard procedure (using the approximation $\Mb=\MS$) as compared
with the use of the extended set, in particular including
the universal relation Love-$\deltaM$. For that purpose we assume as given data a particular set of three quantities, specifically the tidal deformability $\lS$ (or, equivalently, the quadrupolar Love number), the mass $\MS$, and the angular velocity $\OS$. This is a convenient choice, as it can be argued that observations in binary systems and merging neutron stars
can provide such measurements, but the whole argument could proceed in a similar manner with a different initial data set. Second, we show how the numerical accuracy in the extraction of the
parameters of both a polytropic EoS and the MIT bag  EoS (hence, taking into account both neutron stars and quark stars)
improves
when using the extended set of relations.

The paper is arranged as follows: In Section~\ref{sec:review}
we briefly review the quantities involved in the $I$-Love-$Q$-$\deltaM$ relations.
In Section~\ref{sec:correct} we show how to construct a procedure to use the extended set of universal relations sidestepping the approximation $\Mb= \MS$.
Using this procedure, in Section~\ref{sec:errors} we extract the remaining quantities, $\Mb$, $I$ and $Q$, for both a polytropic EoS and the MIT bag EoS for various central pressures and angular velocities. This is also done using the standard procedure
($\Mb= \MS$) in order to compare the relative errors between the two approaches.
Finally, to test the usefulness of the extended universal relations reported in this paper,
we discuss in Section~\ref{sec:infer_eos} a simple model problem to infer EoS parameters, comparing the results with those obtained using the standard universal relations. Section~\ref{sec:conclusions} contains our conclusions along with final remarks on fast-spinning stars in BNS systems. Finally, we devote Appendix \ref{app:fittings}  to discuss the specific case of the $\deltaM$-$I$ and $\deltaM$-$Q$ relations, whose universality had not been previously reported
in the literature.

\section{Review of the quantities}\label{sec:review}
We start reviewing the definitions of the quantities
that describe a rigidly rotating perfect fluid isolated
star in the Hartle-Thorne perturbation
scheme\footnote{We use ``perturbation scheme" to stress the fact that the perturbative
approach entails a choice of class of gauges relative to which the quantities are computed.
We refer to \cite{MRV1} for a precise definition.} \cite{hartle1967,hartle2}, the tidal deformability
in the binary problem (at first order), and the dimensionless counterparts that
enter the universal relations.
It is worth spending first a few lines summarising our notation. We denote as $\NOTATIONS$ any intrinsic quantity $X$ of the star, namely the quadrupolar Love number $\lS$, angular velocity $\OS$, moment of inertia $\IS$, angular momentum $\JS$, spin parameter $\chiS$, quadrupole moment $\QS$ and mass $\MS$. Note that the background quantities $\Mb$ (mass) and $\Rb$ (radius), as well as the second-order contribution of the mass $\deltaM$ are not included in this set. In addition, $\NOTATIONext$ will denote the quantities inferred from the {\it extended} approach.  This set comprises $\miM$, $\miI$ and $\miQ$, together with some related quantities that will be defined where appropriate, such as the relative error $\NOTATIONerrorext$. Similarly, $\NOTATIONstd$ will refer to the  magnitudes of the {\it standard} approach. Finally, when dealing with  numerical computations 
we will use $\NOTATIONH$ to denote any quantity computed semi-analytically to second order in the perturbative approach. Moreover, when fixing the EoS to consider a particular stellar  model, we will employ the notation 
$\NOTATIONmsEOS$. Hence, all the magnitudes derived from this $\NOTATIONmsEOS$ will inherit the same notation, i.e. $\NOTATIONms:=\NOTATIONH(\NOTATIONmsEOS)$.

Let us begin with the isolated star problem. Given an EoS, 
the whole model up to second order exists and
is uniquely determined by the central pressure $P_c$
and the angular velocity of the star $\OS$ as measured by
the observer at rest at infinity (see \cite{MRV2} for the full proof)\footnote{The
fact that the central pressure is gauge independent within
the usual perturbation schemes is shown in \cite{MRV2}.
Let us also stress that although equatorial symmetry has usually been assumed in the literature,
that assumption is not necessary, as shown in \cite{MRV2}.}.
In fact, the scalability property of the perturbations
allows the choice of  $\OS$ as 
the perturbation
parameter, and such that the first and second order
contributions to the corresponding magnitudes consist of a factor
$\OS$ and $\OS^2$, at first and second order respectively, times functions
that only depend on $P_c$ and the EoS (in the remainder
of this section we take the EoS as given, so we do not include the dependence on the EoS explicitly).

In particular, the mass of the rotating star $\MS(P_c, \OS)$
is found to be
\begin{align}\label{eq:mass}
    \MS(P_c, \OS)=\Mb(P_c)+\OS^2 \deltaM(P_c),
\end{align}
where the contribution from the spherically symmetric background configuration $\Mb(P_c)$
is provided by solving the TOV equations
in the interior, and thus only depends on $P_c$,
while the second order contribution to the mass
involves the integration of the second order problem (the $\ell=0$ sector)
to find a quantity $\deltaM(P_c)$.
The TOV problem provides also the value of the radius
of the background configuration, that we denote by $\Rb(P_c)$.

The solution to the first order problem yields the moment of inertia $I(P_c)$, so that the angular momentum of the star,
which is a first order quantity,
is given by
\begin{align*}
    \JS(P_c,\OS)= \OS I(P_c).
\end{align*}
Since $I(P_c)$ corresponds indeed to the moment of inertia
of the star, we will also use the notation $\IS(=I)$
in what follows to make that explicit.

To finish with the isolated configuration, the $\ell=2$ sector of the
second order problem yields another second order quantity $Q(P_c)$ from where the quadrupole moment of the star 
is obtained as
\begin{equation}
     \QS(P_c,\OS)= \OS^2 Q(P_c). \label{eq:qnew}
\end{equation}

The tidal problem in a binary system is treated perturbatively to first order. This
provides the leading order tidal deformability
that, given the EoS, only depends on $P_c$,
and we denote as $\lS(P_c)$.
This quantity is related to the so-called Love number $k_2(P_c)$ via the formula
\[
\lS(P_c)=\frac{2}{3}k_2(P_c)\left(\frac{c^2}{G} C^{-1}(P_c)\right)^5,
\]
where $C(P_c)$ is the compactness of the star, which at first order is given by $C(P_c)=\Mb(P_c)/\Rb(P_c)$.
As it is customary, 
we will refer to the dimensionless quantity $\lS(P_c)$
as the Love number in the following.

As mentioned in the Introduction,
the $I$-Love-$Q$-$\deltaM$ relations found using the two perturbative approaches (both for the isolated problem and for the tidal problem)
refer to the dimensionless and rotation-independent quantities
\begin{align}
    &\overline{I}(P_c):=\frac{\IS(P_c)}{\Mb^3(P_c)},\label{eq:adimIner1}\\
    &\overline{Q}(P_c):=\frac{\QS(P_c,\OS) \Mb(P_c)}{\JS^2(P_c,\OS)}=\frac{Q(P_c) \Mb(P_c)}{\IS^2(P_c)},\label{eq:adimQ1}\\
    &\overline{\deltaM}(P_c):=\OS^2\deltaM(P_c)\frac{\Mb^3(P_c)}{\JS^2(P_c,\OS)}
    =\deltaM(P_c)\frac{\Mb^3(P_c)}{\IS^2(P_c)},\label{eq:adimdM1}
\end{align}
and $\lS(P_c)$.
The fact that these dimensionless quantities depend on the non-observable
TOV mass $\Mb$ makes the
use of the universal relations and the translation to the star quantities
less straightforward, a priori.
We show
in the following section how to employ the universal relations
involving also $\deltaM$ to infer the value of $\Mb$ and thus enable a precise correspondence  between both sets of quantities. Although we do not employ the universal relations $\deltaM$-$I$ and $\deltaM$-$Q$ in the present work, we include them in Appendix \ref{app:fittings} for completeness.

\section{Application of universal relations: the extended approach}\label{sec:correct}

In a binary neutron star system the dimensionless and leading order tidal deformability $\lS$ of the component stars, as well as their masses $\MS$, can be inferred from the inspiral gravitational-wave signal by correlating the waveform strain against theoretical templates 
(see~\cite{Flanagan2008,De:2018uhw,LIGOScientific:2018hze}). 
The spins of merging neutron stars are also imprinted on the gravitational waveform, primarily on a quantity called the effective spin parameter $\chi_{\rm eff}$ (see e.g.~\cite{Zhu:2018}). Its measurement may be used to statistically infer the individual spin components (and the angular velocity $\OS$). The observation of GW170817 yielded values consistent with zero effective spin~\cite{GW170817-properties}. As future gravitational-wave (GW) detectors increase the number of binary neutron star merger observations, the shape of the $\chi_{\rm eff}$ distribution might be constrained.

Thus, we take the set $(\lS, \MS, \OS)$
as known (observed) quantities 
from where to obtain the value of the remaining stellar quantities.
To be precise, our aim is to extract
the quantities $\IS$ and $\QS$ 
in terms of some given values for
$(\lS, \MS, \OS)$
using the universal relations that relate $\overline{I}$, $\lS$, $\overline{Q}$, and $\overline{\deltaM}$.
For the sake of clarity, in the following we refer to
this set of $I$-Love-$Q$-$\deltaM$ 
relations (dropping the bars) as the extended (``$\ext$'') set of relations, in contrast to the standard (``$\std$'') 
$I$-Love-$Q$ set. We will also use ``extended"
for any procedure that uses the extended set to obtain $\Mb$,
and ``standard" for the usual approach that assumes $\Mb=\MS$. Let us stress that one could take any other trio of variables as given data, and the analysis would follow a similar procedure.

We start by writing down the convenient expressions
for $\IS$ and $\QS$ (we drop the dependence on $P_c$ and $\OS$ not to overwhelm the text).
First, from Eq.~\eqref{eq:adimIner1} we have
\begin{align}
    \IS=\Mb^3\,\overline{I},\label{eq:Ibona}
\end{align}
while a combination of \eqref{eq:qnew} and \eqref{eq:adimIner1} with \eqref{eq:adimQ1}
allows us to write
\begin{align}
    \QS=\OS^2\Mb^5\,\overline{I}\,{}^2\,\overline{Q}.\label{eq:Qbona}
\end{align}
To obtain $\overline{I}$ and $\overline{Q}$
from the universal relations, we need $\Mb$.
For that, we can exploit the universal relations involving
$\overline{\deltaM}$ while making use of
\begin{align}
    \overline{\deltaM}=\frac{\MS-\Mb}{\OS^2\Mb^3\,\overline{I}\,{}^2 },\label{eq:adimdM2}
\end{align}
that follows from \eqref{eq:adimdM1} in combination with 
\eqref{eq:mass} and \eqref{eq:adimIner1}.

The individual fitting formulae inherited from the universality of the relations can be expressed as
\begin{align}
    \ln \indi=\pol_{\indi}(\ln\lS),\label{eq:fittingcurve}
\end{align}
where $\indi$ stands for any element in the set $\{\overline{I},\overline{Q},\overline{\deltaM}\}$, and
\begin{align*}
    \pol_{\indi}\left(x\right)&=a_{\indi} + b_{\indi} x + c_{\indi} x^2 + d_{\indi} x^3 + e_{\indi}x^4,
\end{align*}
with values for $\{a_{\indi},b_{\indi},c_{\indi},d_{\indi},e_{\indi}\}$ given in Table \ref{tab:values_fitting}.

\begin{table}[t!]
\setlength{\tabcolsep}{2.1pt}
\renewcommand{\arraystretch}{1.5}
    \centering
    \begin{tabular*}{\linewidth}{@{\extracolsep{\fill}} c c c c c c }
    \hline\hline
         $\indi$ & $a_\indi$ & $b_\indi$ & $c_\indi$ & $d_\indi$ & $e_\indi$ \\
    \hline
         $\overline{I}$  & $1.47$  & $0.0817$ & $0.0149$ & $2.87\times10^{-4}$ & $-3.64\times10^{-5}$ \\
         $\overline{Q}$ & $0.194$ & $0.0936$ & $0.0474$ & $-4.21\times10^{-3}$ & $1.23\times10^{-4}$ \\
         $\overline{\deltaM}$ & $-1.619$ & $0.255$ & $-0.0195$ & $-1.08\times10^{-4}$ & $1.81\times10^{-5}$ \\
    \hline\hline
    \end{tabular*}
    \caption{Parameters for the fitting curve of Eq.~(\ref{eq:fittingcurve}). The first two rows have been obtained from \cite{Yagi:2013awa}, whereas the last row is from \cite{reina-sanchis-vera-font2017} after applying a suitable logarithm base conversion.}
    \label{tab:values_fitting}
\end{table}

For a given value of $\lS$, the extraction of
$\overline{I}$ and $\overline{Q}$ is direct
from the standard $I$-Love-$Q$ relations
\begin{align}
&\ln\overline{I}=\pol_{\overline{I}}\left(\ln\lS\right),\label{eq:I-Love}\\
&\ln\overline{Q}=\pol_{\overline{Q}}\left(\ln\lS\right).\label{eq:Q-Love}
\end{align}
To extract
the value of $\Mb$ we use
the universal relation found in \cite{reina-sanchis-vera-font2017} for
Love-$\deltaM$, which has the form $\ln\overline{\deltaM}=\pol_{\overline{\deltaM}}(\ln\lS)$. Using \eqref{eq:adimdM2} we obtain
\begin{align}
\ln\frac{\MS-\Mb}{\OS^2\Mb^3\,\overline{I}\,{}^2}=\pol_{\overline{\deltaM}}(\ln\lS).\label{eq:dM-Love}
\end{align}
Since the left-hand-side of \eqref{eq:dM-Love} is monotonically decreasing,
given a value of $\lS$ this equation provides a unique value of $\Mb$,
which we call $\miM$. 
As a result, using \eqref{eq:Ibona}-\eqref{eq:Qbona},
the values of $\IS$ and $\QS$ extracted from the extended universal relations,
that we denote by $\miI$ and $\miQ$ respectively, read
\begin{align}
    &\miI=(\miM)^3\,\overline{I},\label{eq:mi_I}\\
    &\miQ=\OS^2(\miM)^5\,\overline{I}\,{}^2\,\overline{Q}.\label{eq:mi_Q}
\end{align}

In contrast to the above procedure, 
the standard approach assumes $\Mb=\MS(=\suM)$, and the extracted values of $\IS$ and $\QS$,
which we denote $\suI$ and $\suQ$ respectively, reduce to
\begin{align}
    &\suI=\MS^3\,\overline{I},\label{eq:su_I}\\
    &\suQ=\OS^2\MS^5 \,\overline{I}\,{}^2\,\overline{Q}.\label{eq:su_Q}
\end{align}

The magnitude of the systematic errors in the use of the universal relations arising from the approximation $\Mb=\MS$ depends on the rotational velocity of the star. In the following section we assess such errors
by comparing the results obtained in the extraction of $I_{S}$ and $Q_{S}$ using the above extended procedure with those obtained
following the standard approach.

\section{Numerical relevance of the  computation}\label{sec:errors}

Our procedure considers a data set which is obtained by solving the perturbative model computationally. This data set is assumed to describe potential observations of neutron star properties. 
First, for a given EoS, we consider different values of $P_c$ and $\OS$ as initial data to solve the perturbative models numerically
(both for the isolated and tidal problems).
We denote respectively by $\MbH(P_c)$, $\RbH(P_c)$, $\ISH(P_c)$, $\dMH(P_c)$, $\QH(P_c)$ and $\lH(P_c)$
the values of $\Mb(P_c)$, $\Rb(P_c)$, $\IS(P_c)$, $\deltaM(P_c)$, $Q(P_c)$, and $\lS(P_c)$
computed numerically.
We also obtain $\QSH(P_c,\OS)=\OS^2\QH(P_c)$.

Then, considering also $\MSH(P_c,\OS)=\MbH(P_c)+\OS^2\dMH(P_c)$ we construct the trio
$(\lH,\MSH,\OS)$ and extract, using the
full set of $I$-Love-$Q$-$\deltaM$ relations, the values of
$\miM$, $\miI$ and $\miQ$
as explained in Sec.~\ref{sec:correct}. We repeat the process employing the standard method to extract $\suI$ and
$\suQ$, i.e. using only
the $I$-Love-$Q$ relations with $\suM=\MS$.

To calculate the deviations between the two approaches we define the relative errors for each $P_c$ as
\begin{align}
    &\mierrorM:=\frac{|\miM-\MbH|}{\MbH},\label{eq:mi_error_M0}\\
    &\mierrorI:=\frac{|\miI-\ISH|}{\ISH},\label{eq:mi_error_I}\\
    &\mierrorQ:=\frac{|\miQ-\QSH|}{\QSH},\label{eq:mi_error_Q}
\end{align}
and
\begin{align}
    &\suerrorM:=\frac{|\MSH-\MbH|}{\MbH}=\OS^2\frac{|\dMH|}{\MbH},\label{eq:su_error_M0}\\
    &\suerrorI:=\frac{|\suI-\ISH|}{\ISH},\label{eq:su_error_I}\\
    &\suerrorQ:=\frac{|\suQ-\QSH|}{\QSH}\label{eq:su_error_Q},
\end{align}
respectively. Let us stress that, for a given EoS, all six quantities depend on
$P_c$ and $\OS$ by construction. Observe also that $\suerrorM$ accounts
for the ratio between the contribution of the mass at second order
over the TOV mass of the background configuration. The quantity
$\suerrorM$ thus provides
the basic measure of the departure of the extended procedure with respect
to the standard one. This departure propagates to the extraction of the moment of inertia and of the quadrupole moment
using the standard approach, yielding deviations in these two quantities.
On the other hand, the quantity $\mierrorM$ gives a measure of the precision
of the extraction of $\Mb$ from the particular universal relation
Love-$\deltaM$. 

To cover a wider range of possible scenarios, we work with two different EoSs, namely a relativistic polytrope and the MIT bag model. The former has a vanishing energy density at the boundary of the star, whereas the later presents a jump there.

\begin{figure*}[htb!]
    \centering
    \includegraphics[width=\textwidth]{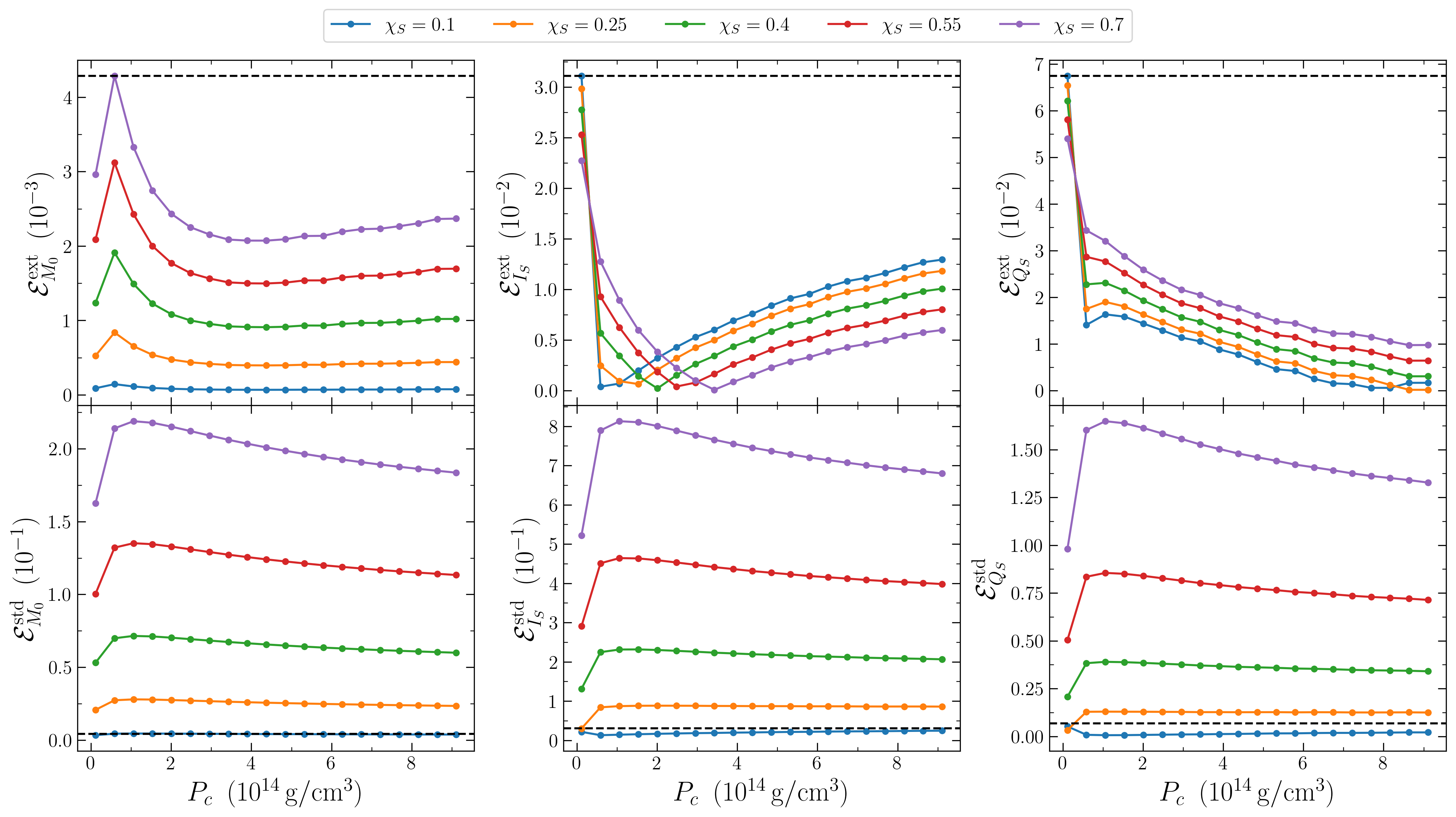}
    \caption{Relative errors of the inferred quantities $\Mb$, $\IS$ and $\QS$ (from left to right) for the polytropic EoS \eqref{eq:polytropic_EoS} as a function of the central pressure $P_c$ and for different values of the spin parameter $\chiS$. The upper panel of each column shows the relative errors obtained when using the extended procedure \eqref{eq:mi_error_M0}-\eqref{eq:mi_error_Q} and the lower one shows the errors produced by the standard procedure \eqref{eq:su_error_M0}-\eqref{eq:su_error_Q}. To facilitate the comparison we have included horizontal dashed black lines at each column indicating the highest value of the corresponding errors of the extended approach.}
    \label{fig:plots_polytropic}
\end{figure*}

\subsection{Relativistic polytropic EoS}
One of the simplest models to describe relativistic stars
is given by a polytropic EoS, for which the pressure $P$ and density $\rho$ are related by a power law
\begin{align}
    P=K\rho^\gamma,\label{eq:polytropic_EoS}
\end{align}
where $K$ is constant
and $\gamma$ is related to the polytropic index $n$ via $\gamma:=1+1/n$. The energy density is then given by
\begin{align}
    E=\rho+\frac{P}{\gamma-1}.
\end{align}

In dimensionless geometrised units $(c=G=M_\odot=1)$ we set $K=100$ and $\gamma=2$ as in \cite{Font:1999}, with a central density ranging from $\rho_c=4.27\times10^{-4}$ to $\rho_c=3.84\times10^{-3}$. This corresponds to a range in the central pressure from $P_c=1.12\times10^{13}$ \si{\g/\cm^{3}} to $P_c=9.11\times10^{14}$ \si{\g/\cm^{3}}. The smallest and largest TOV masses in the sample are given by $\Mb=0.77\,M_\odot$ and $\Mb=1.64\,M_\odot$, respectively.

Figure~\ref{fig:plots_polytropic} shows the relative errors of $M_0$, $I_{S}$ and $Q_S$,  defined in \eqref{eq:mi_error_M0}-\eqref{eq:su_error_Q}, as functions of the central pressure $P_c$ for different values of the spin
parameter\footnote{
In units of $c=G=1$, and to second order in perturbation theory, the dimensionless spin parameter is given by $\chiS=\IS\OS/\Mb^2$.},
$\chiS$. In general 
we observe that the extended procedure yields more accurate predictions (smaller errors) than the standard counterpart. 

\subsection{MIT bag EoS}
The internal structure of quark stars is usually described with the so-called MIT bag model \cite{chodos_1:1974,chodos_2:1974,degrand:1975}, for which the pressure $P$ and the energy density $E$ are related by
\begin{align}
    P = \frac{1}{3}(E - 4 B),\label{eq:mitbag_EoS}
\end{align}
where $B\leq E/4$ is the \textit{bag constant}, accounting for the confinement of quarks. It is straightforward to check that the vanishing of the pressure (which determines the boundary of the star) implies a non-vanishing energy density at the surface $E(\Rb)=4B$. In our numerical calculations we set the bag constant to $B=10^{14}$~\si{\g/\cm^{3}} 
as in \cite{Colpi:1992}, and we consider a number of values for the central energy density ranging from $E_c = 6\times10^{14}$~\si{\g/\cm^{3}} to $E_c=4.4\times10^{15}$~\si{\g/\cm^{3}}, which correspond to $P_c=6.63\times10^{13}$~\si{\g/\cm^{3}} and $P_c=1.33\times10^{15}$~\si{\g/\cm^{3}}, respectively. The non-rotating masses of these stellar models range between $\Mb=1.13\,M_\odot$ and $\Mb=2.02\,M_\odot$.

\begin{figure*}[t]
    \centering
    \includegraphics[width=\textwidth]{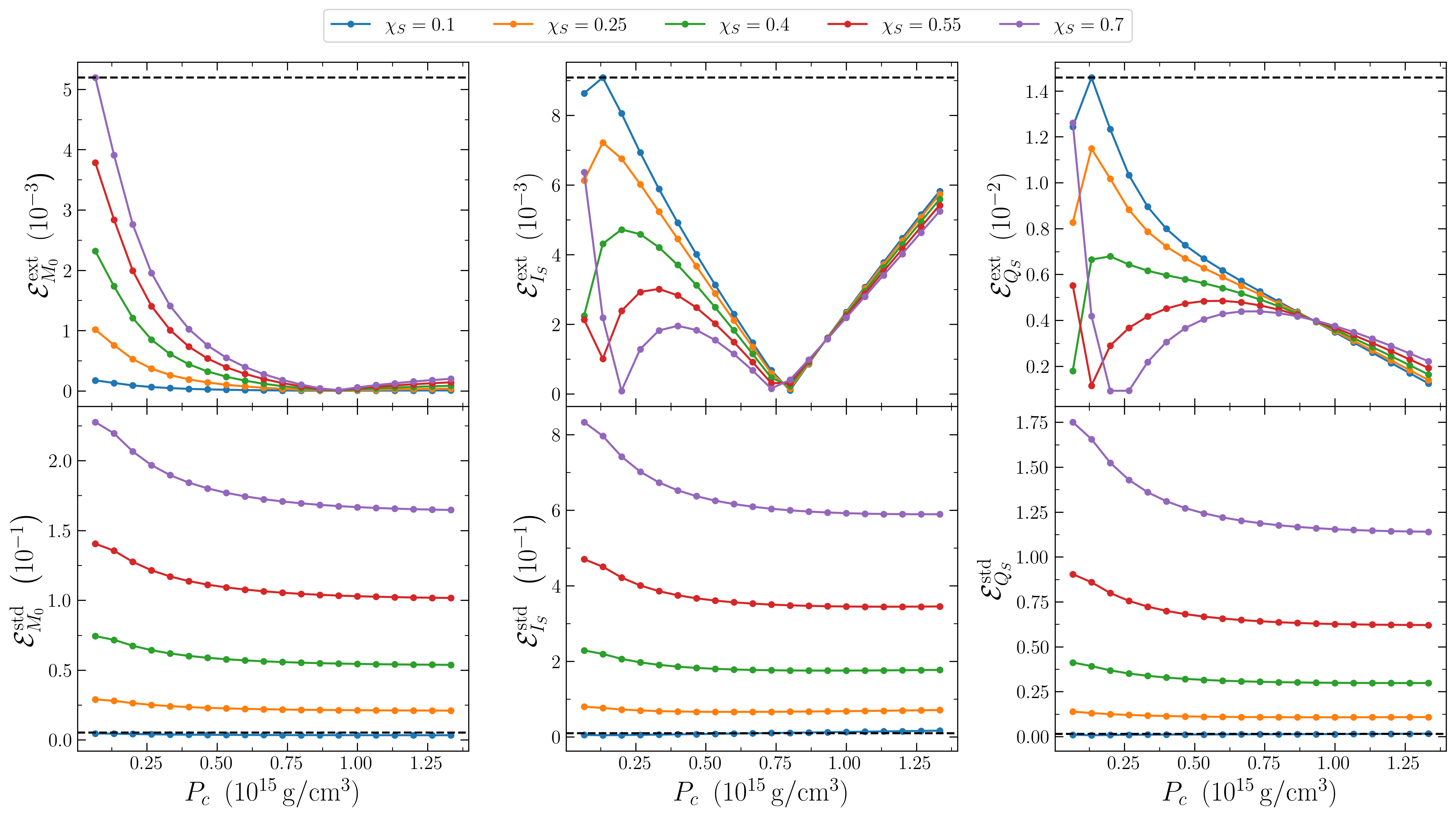}
    \caption{Relative errors of the inferred quantities $\Mb$, $\IS$ and $\QS$ (from left to right) for the MIT bag EoS \eqref{eq:mitbag_EoS} as a function of the central pressure $P_c$ and for different values of the spin parameter $\chiS$. 
    The upper panel of each column shows the relative errors obtained when using the extended procedure \eqref{eq:mi_error_M0}-\eqref{eq:mi_error_Q} and the lower panel shows the errors produced by the standard procedure \eqref{eq:su_error_M0}-\eqref{eq:su_error_Q}. As in Fig.~\ref{fig:plots_polytropic}, the horizontal dashed black lines in each column indicate the highest value of the corresponding errors of the extended approach.}
    \label{fig:plots_MITBAG}
\end{figure*}

As done for the relativistic polytropic EoS in Fig.~\ref{fig:plots_polytropic}, we display in Fig.~\ref{fig:plots_MITBAG} the relative errors defined in \eqref{eq:mi_error_M0}-\eqref{eq:su_error_Q} as functions of $P_c$ for different values of $\chiS$ for the MIT bag EoS. Again, the standard approach is shown to yield less accurate predictions than the extended procedure.

\section{Inference of EoS parameters}\label{sec:infer_eos}

We turn now to quantify in a simple setting the usefulness of the universal relations involving $\deltaM$ in the inference of EoS parameteres. To do so we consider two different model problems which we will use to infer parameters of the EoS employing the universal relations both in the standard and in the extended approaches. In the first case we assume that the observed star is governed by a polytropic EoS with yet unknown values of $K$, $\gamma$, and $P_c$; for the second case we consider the MIT bag EoS, for which the procedure to infer the values of $B$ and $P_c$ follows similarly.

The core of the idea in the first problem 
is to construct a polytropic stellar model
defined by a set of ten compatible quantities
$\{\Kms,\gammams,\Pcms;\Mbms,\Rbms;\Mms,\lms,\Ims,\Qms,\Oms\}$
using the perturbative analytical models.
To do that one prescribes some values for the trio $(\Kms,\gammams,\Pcms)$ and computes $\{\Mbms,\Rbms;\Mms,\lms,\Ims,\Qms\}$ numerically
for some choice of the angular velocity $\Oms$. The exercise consists in determining how the standard universal relations compare with the extended universal relations in resolving the EoS parameters.
As a first step, we fix the value of $\Kms$ and
only vary $\gamma$ and $P_c$.
This simple two-dimensional exercise,
discussed in subsection \ref{sec:given_K},
is used to illustrate the methodology of the inference. This helps to understand the more general setup presented in subsection \ref{sec:free_K}, where $\Kms$ is not fixed.
The second problem is discussed in Sec.~\ref{sec:mit_bag}, where the approach of Sec.~\ref{sec:given_K} will be applied to the MIT bag EoS to infer $B$ and $P_c$.

\subsection{Polytropic EoS: Inferring $\gamma$ and $P_c$ with fixed $K$}\label{sec:given_K}

\begin{figure*}[htb]
\centering
    \includegraphics[width=\linewidth]{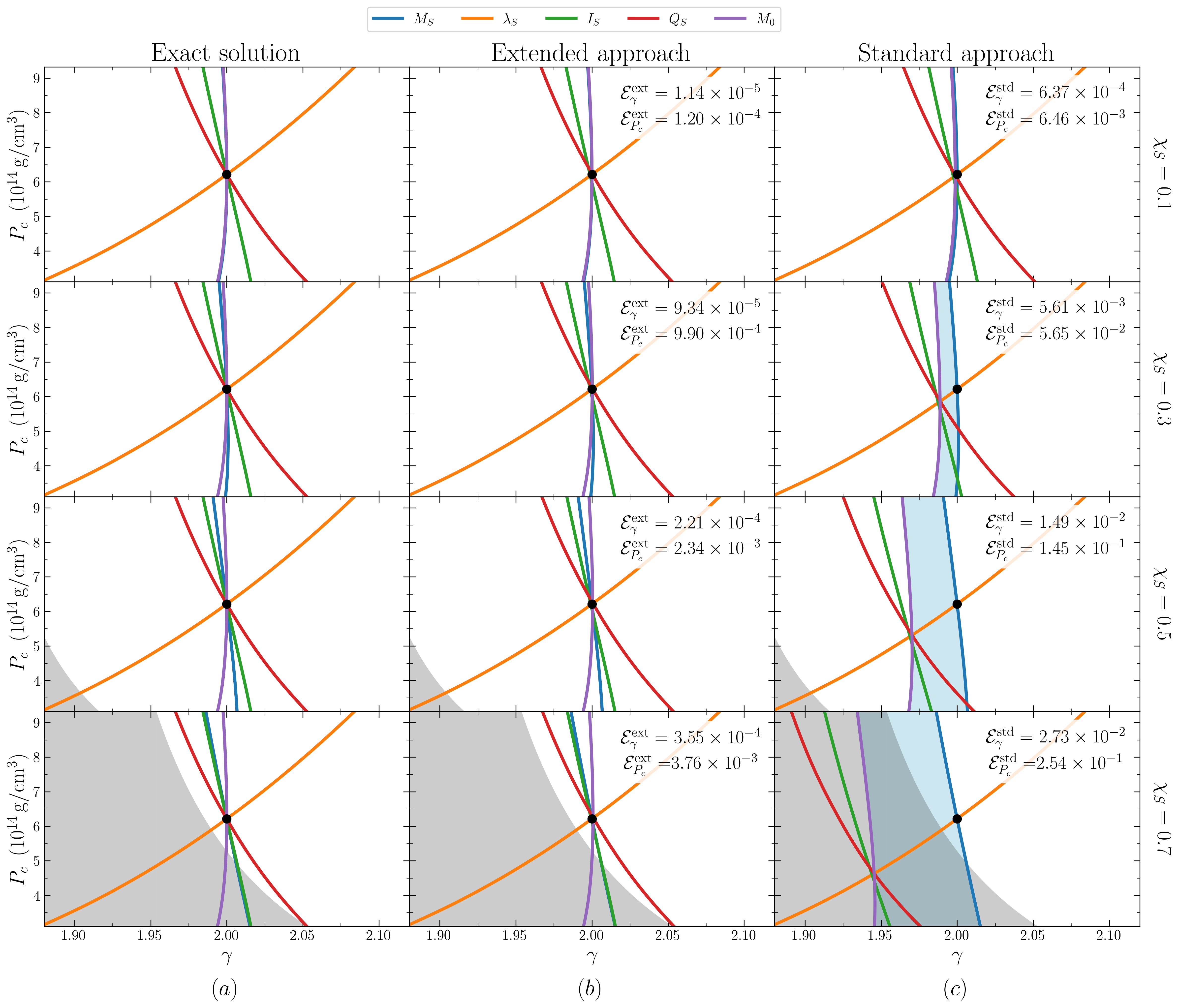}
\caption{The curves $\Clove$ and $\CMS$,
defined in \eqref{eq:curves_exact}, together with
(a) $\CI$, $\CQ$, $\CMb$ from \eqref{eq:curves_exact_2},
(b) $\CIR$, $\CQR$, $\CMbR$ from \eqref{eq:curves_good} and 
(c) $\CIW$, $\CQW$, $\CMbW$ from \eqref{eq:curves_bad}, for different dimensionless spin parameters $\chiS$. We have considered a stellar model with $\Kms=100$, $\gammams=2$ and $\Pcms=6.22\times10^{14}\,\mathrm{g}/\mathrm{cm}^3$. The grid has been constructed for the values $\gamma\in[1.88, 2.12]$ and $P_c\in[3.11,9.32]\times10^{14}\,\mathrm{g}/\mathrm{cm}^3$. The grey shaded regions represent the points ($\gamma, P_c$) for which $\OKH(\gamma,P_c)<\Oms$, and the blue shaded regions in column (c) illustrate the indeterminacy inherited from the assumption $\suM=\Mms$. The legend in columns (b) and (c) include the relative errors defined in \eqref{eq:error_gamma_R}-\eqref{eq:error_Pc_R} and \eqref{eq:error_gamma_W}-\eqref{eq:error_Pc_W}, respectively.}
\label{fig:poly_K_fixed}
\end{figure*}

The computational construction of our first  exercise is described as follows. For a fixed $\Kms=100$, we consider a discrete range of values for the unknowns $\gamma$ and $P_c$ and construct a grid to account for all possible combinations\footnote{The units of $K$ in the SI depend on $\gamma$ via $[K]=\frac{\si{kg}}{\si{m}\cdot \si{s}^2}\left(\frac{\si{m}^3}{\si{kg}}\right)^\gamma$, so we will always write $K$ in units of $c=G=M_\odot=1$, where it is dimensionless.}. We then use the perturbative analytical 
model to compute numerically
the quantities $\MbH(\gamma,P_c)$, $\RbH(\gamma,P_c)$, together with
$\dMH(\gamma,P_c)$, $\lH(\gamma,P_c)$,
$\ISH(\gamma,P_c)$, $\QH(\gamma,P_c)$  for each pair of values $(\gamma,P_c)$ in the grid.
Using the background quantities we compute the Kepler limit of the angular velocity $\OKH(\gamma,P_c)$,
\[
\OKH(\gamma, P_c):=\sqrt{\frac{\MbH(\gamma, P_c)}{\RbH^3(\gamma, P_c)}},
\]
at each cell.
Then, we interpolate the values of the grid to represent each quantity
as a smooth surface over the plane $\{\gamma,P_c\}$. The results are displayed in Fig.~\ref{fig:poly_K_fixed}.

For a given value of the angular velocity $\Oms$, we
construct the surfaces for $\MSH(\gamma,P_c,\Oms)=\MbH(\gamma,P_c)+\Oms{}^2\dMH(\gamma,P_c)$ and $\QSH(\gamma,P_c,\Oms)=\Oms{}^2 \QH(\gamma,P_c)$
over the plane  $\{\gamma,P_c\}$.
Due to the existence of a maximal angular velocity $\OKH$, there might be some points $(\gamma,P_c)$ for which $\OKH(\gamma,P_c)<\Oms$; that is, stellar configurations rotating faster than physically permitted.
These regions have been indicated
by a grey shaded area in Fig.~\ref{fig:poly_K_fixed}.

Once the whole plane has been dressed with the
interpolated surfaces defined by
the quantities $\{\MbH,\MSH,\lH,\ISH,\QSH\}$,
we set our stellar model by picking a
point $(\gammams, \Pcms)$.
The values of the quantities in the set $\{\MbH,\MSH,\lH,\ISH,\QSH\}$ evaluated on $(\gammams, \Pcms)$,
and using $\Oms$,
yield the quantities for our stellar model
$\{\Mbms,\Mms,\lms,\Ims,\Qms\}$,
that is $\Mbms=\MbH(\gammams,\Pcms)$, $\Mms=\MSH(\gammams,\Pcms,\Oms)$ and so on.
Now, for each quantity in the set
$\{\Mbms,\Mms,\lms,\Ims,\Qms\}$, one can draw the intersection
of the plane determined by its value and the
corresponding interpolated surface
to obtain a curve $\Curve$ on the plane $\{\gamma,P_c\}$, specifically
\begin{align}
    &\Clove:=\{\lH(\gamma,P_c)=\lms\},\nonumber\\
    &\CMS:=\{\MSH(\gamma,P_c,\Oms)=\Mms\},\label{eq:curves_exact}
\end{align}
and
\begin{align}
    &\CI:=\{\ISH(\gamma,P_c)=\Ims\},\nonumber\\
    &\CQ:=\{\QSH(\gamma,P_c,\Oms)=\Qms\},\label{eq:curves_exact_2}\\
    &\CMb:=\{\MbH(\gamma,P_c)=\Mbms\}.\nonumber
\end{align}
For instance, $\Clove$ represents all the points $(\gamma,P_c)$
with the same value of $\lms$.
Clearly, all curves must intersect at the point
describing the stellar model by construction\footnote{Any deviation should be caused by the numerical precision used for the interpolation.} [see Fig.~\ref{fig:poly_K_fixed} (a)].

To see how well the use of the extracted quantities
from the universal relations fares in the determination
of $\gamma$ and $P_c$, we proceed as follows.
First, we pick the values $\lms$, $\Mms$  and $\Oms$,
and extract the corresponding $\miM$, $\miI$, and $\miQ$
using the extended set of universal relations 
\eqref{eq:I-Love}-\eqref{eq:dM-Love},
together with \eqref{eq:mi_I} and \eqref{eq:mi_Q}.
We now proceed to obtain the curves determined
by the intersection of the interpolated surfaces
and the corresponding planes defined by the values
$\{\lms,\Mms,\miI,\miQ,\miM\}$. Explicitly, the curves
correspond to the set~\eqref{eq:curves_exact} plus
\begin{align}
    &\CIR:=\{\ISH(\gamma,P_c)=\miI\},\nonumber\\
    &\CQR:=\{\QSH(\gamma,P_c,\Oms)=\miQ\},\label{eq:curves_good}\\
    &\CMbR:=\{\MbH(\gamma,P_c)=\miM\}.\nonumber
\end{align}
These five curves are plotted in column (b) of 
Fig.~\ref{fig:poly_K_fixed} for different values of $\chiS$. Ideally, if the extracted quantities were correct, all curves would coincide at the stellar model $(\gammams,\Pcms)$, and thus we would recover the results plotted in column (a). The corresponding relative errors, of ${\cal O}(10^{-3})$ at most, are displayed in the legends of the figure (specific details on the computation of these errors are reported below).

\begin{table*}[t]
\begin{tabular}{ccccccccccc}
\hline
Model & $E_c$ & $P_c$ & $\Mb$ & $\Rb$ & $\mierrorgamma$ & $\suerrorgamma$ & $\mierrorPc$ & $\suerrorPc$ \\
 & $(\times10^{-3})$ & $(\times10^{-4})$ &  &  & $(\times10^{-4})$ & $(\times10^{-4})$ & $(\times10^{-4})$ & $(\times10^{-4})$\\
\hline
AU0 & 1.444 & 1.639 & 1.400 & 9.585 & 6 & 333 & 79 & 3764 \\

AU1 & 1.300 & 1.356 & 1.352 & 9.768 &  6 & 336 & 89 & 3853 \\

AU2 & 1.187 & 1.149 & 1.308 & 9.922 & 7 & 339 & 97 & 3918 \\

AU3 & 1.074 & 0.957 & 1.257 & 10.086 & 7 & 343 & 106 & 4006 \\

AU4 & 0.961 & 0.780 & 1.198 & 10.262 & 8 & 346 & 114 & 4096 \\

AU5 & 0.863 & 0.639 & 1.139 & 10.426 & 8 & 350 & 118 & 4182 \\
\hline
\end{tabular}
\caption{Polytropic AU0-AU5 models from \cite{Stergioulas:2003ep,Dimmelmeier:2005zk} rotating at their maximum angular velocity $\OKH(P_c)$.}
\label{tab:AU}
\end{table*}

To assess how this 
difference compares to the standard use 
of the universal relations,  
we repeat the exercise using the standard procedure, 
considering $\suM=\Mms$. Explicitly, taking the values $\lms$, $\Mms$, and $\Oms$, we use 
\eqref{eq:I-Love}-\eqref{eq:Q-Love},
together with \eqref{eq:su_I}-\eqref{eq:su_Q},
to extract $\suI$ and $\suQ$. The five curves
are thus given by \eqref{eq:curves_exact} plus
\begin{align}
    &\CIW:=\{\ISH(\gamma,P_c)=\suI\},\nonumber\\
    &\CQW:=\{\QSH(\gamma,P_c,\Oms)=\suQ\},\label{eq:curves_bad}\\
    &\CMbW:=\{\MbH(\gamma,P_c)=\Mms\}.\nonumber
\end{align}
The results are plotted in column (c) of 
Fig.~\ref{fig:poly_K_fixed}.
Notice that the use of $\suM=\Mms$ not only increases the inaccuracy in the computation of $\gamma$ and $P_c$ (to maximum values of ${\cal O}(10^{-1})$) but also 
introduces significant uncertainty in the procedure, which we indicate by the blue shaded region in Fig.~\ref{fig:poly_K_fixed} (c). This uncertainty, as expected, increases with $\chiS$.

Since the difference between both approaches lies on the proper determination of the value of $\Mb$, we compare next the errors
introduced by the extended and the standard approaches
using the pair of curves $\Clove$ and $\CMbR$ in the former case,
and the pair of curves $\Clove$ and $\CMbW$ in the latter.
Specifically, we define the relative errors in the computation
of $\gamma$ and $P_c$ as
\begin{align}
    &\mierrorgamma := \frac{\lvert\gamma(\Clove\cap\CMbR)-\gammams\lvert}{\gammams},\label{eq:error_gamma_R}\\
    &\mierrorPc := \frac{\lvert P_c(\Clove\cap\CMbR)-\Pcms\lvert}{\Pcms},\label{eq:error_Pc_R}
\end{align}
for the extended approach, and
\begin{align}
    &\suerrorgamma := \frac{\lvert\gamma(\Clove\cap\CMbW)-\gammams\lvert}{\gammams},\label{eq:error_gamma_W}\\
    &\suerrorPc := \frac{\lvert P_c(\Clove\cap\CMbW)-\Pcms\lvert}{\Pcms},\label{eq:error_Pc_W}
\end{align}
for the standard counterpart. The numerical results for the relative errors
are displayed in Fig.~\ref{fig:poly_K_fixed} (b) and (c), respectively. The relative errors in the inference of $\gamma$ in the standard approach are approximately $56-77$ times larger than those in the extended counterpart, and for $P_c$ they are about $54-68$ times larger.
From these results, and taking into account that the approximation $\suM=\Mms$ is equivalent to $\Oms^2\deltaM=0$, it follows that the differences between the corresponding errors in both approaches must increase for higher values of $\Oms$, and thus of $\chims$.

To complete this study we consider a wider set of stellar configurations and include in Table~\ref{tab:AU} the relative
errors $\mierrorgamma$, $\suerrorgamma$ and $\mierrorPc$, $\suerrorPc$ for the AU0-AU5 polytropic models proposed in \cite{Stergioulas:2003ep,Dimmelmeier:2005zk}, but taking $\Oms=\OKH(\gammams,\Pcms)$ in our case (i.e.~all models rotate at their maximum allowed angular velocity). As expected, for these models we also obtain
$\suerrorgamma>\mierrorgamma$ and $\suerrorPc>\mierrorPc$, with a systematic difference of more than one order of magnitude in favour of the extended approach. More precisely, the deviations of the standard approach are around $44-58$ times higher for $\gamma$ and $35-48$ times higher for $P_c$.

In this particular two-dimensional exercise one could actually determine the value of $(\gammams, \Pcms)$ from the intersection of $\Clove$ and $\CMS$ alone (as long as there is no degeneracy)
\begin{align}
    (\gammams,\Pcms)=\Clove\cap\CMS,
\end{align}
without resorting to the universal relations.
However, if we no longer fix $\Kms$ but rather take it as an unknown, the trio $(\Kms,\gammams, \Pcms)$ is not uniquely determined by $\Clove$ and $\CMS$ alone. Moreover, in this case the choice of the approach employed for the universal relations (extended vs standard) makes a major difference in the inference of $K$, $\gamma$ and $P_c$, 
as we show next.

\begin{figure*}[t]
\centering
\begin{tabular}{ccc}
    \includegraphics[width=\linewidth]{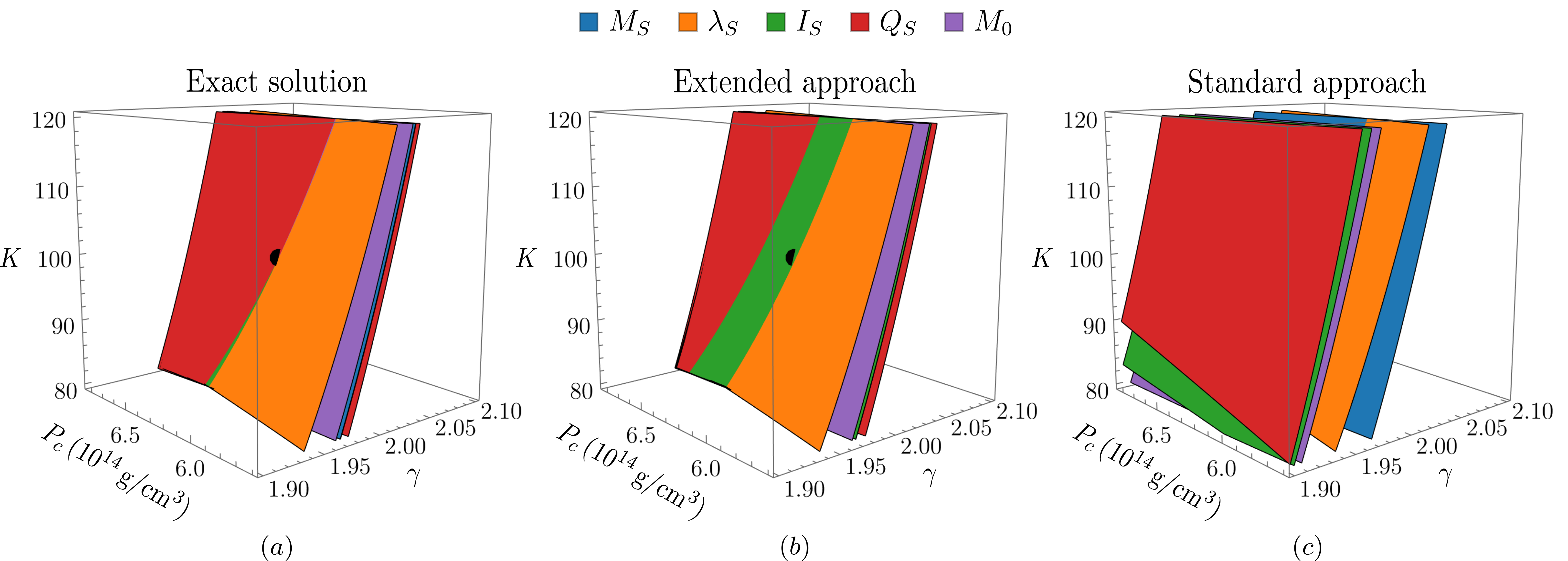}
\end{tabular}
\caption{$\Slove$ and $\SMS$ defined by \eqref{eq:planes_exact} together with (a) $\SI$, $\SQ$, $\SMb$ from \eqref{eq:planes_exact_2}, (b) $\SIR$, $\SQR$, $\SMbR$ from \eqref{eq:planes_good} and (c) $\SIW$, $\SQW$, $\SMbW$ from \eqref{eq:planes_bad} for fixed dimensionless spin parameter, $\chiS=0.7$. We use the same stellar model $(\Kms, \gammams, \Pcms)$ as in Fig.~\ref{fig:poly_K_fixed}. We construct the grid for the triad $(K, \gamma, P_c)$ with the ranges
$K\in[80, 120]$, 
$\gamma\in[1.9,2.1]$ and $P_c\in[5.59,6.84]\times10^{14}\,\mathrm{g}/\mathrm{cm}^3$. We omit the grey and blue shaded regions as represented in Fig.~\ref{fig:poly_K_fixed} for visual clarity.
}
\label{fig:grid_3d}
\end{figure*}

\subsection{Polytropic EoS: Inferring $K$, $\gamma$ and $P_c$}\label{sec:free_K}

We now generalize the previous analysis
by letting the polytropic constant $K$ run within a certain range. For a given $\Oms$ we compute
$\lH(K,\gamma,P_c)$, $\MSH(K,\gamma,P_c,\Oms)$, 
$\ISH(K,\gamma,P_c)$, $\QSH(K,\gamma,P_c,\Oms)$ and $\MbH(K,\gamma,P_c)$,
together with the Kepler limit $\OKH(K,\gamma,P_c)$
in every cell of our three-dimensional grid,
and we interpolate to find the quantities as functions
over the space $\{K,\gamma,P_c\}$.

To fix the stellar model we now select a point $\pms:=(\Kms,\gammams,\Pcms)$.
Evaluating
$\{\lH,\MSH,\ISH,\QSH,\MbH\}$
on $\pms$ we obtain
the stellar model
quantities $\{\lms,\Mms,\Ims,\Qms,\Mbms\}$.
In this case, the expressions 
\begin{align}
    &\Slove:=\{\lH(K,\gamma,P_c)=\lms\},\nonumber\\
    &\SMS:=\{\MSH(K,\gamma,P_c,\Oms)=\Mms\},\label{eq:planes_exact}
\end{align}
and
\begin{align}
    &\SI:=\{\ISH(K,\gamma,P_c)=\Ims\},\nonumber\\
    &\SQ:=\{\QSH(K,\gamma,P_c,\Oms)=\Qms\},\label{eq:planes_exact_2}\\
    &\SMb:=\{\MbH(K,\gamma,P_c)=\Mbms\},\nonumber
\end{align}
determine the corresponding surfaces in the $\{K,\gamma,P_c\}$-space
for each quantity in the set $\{\lms,\Mms,\Ims,\Qms,\Mbms\}$ which, 
by construction, will coincide at least on $\pms$.

As before, we first use
the quantities extracted from the extended set of
universal relations $\miI$, $\miQ$, $\miM$, and construct
the corresponding surfaces, determined by
\begin{align}
    &\SIR:=\{\ISH(K,\gamma,P_c)=\miI\},\nonumber\\
    &\SQR:=\{\QSH(K,\gamma,P_c,\Oms)=\miQ\},\label{eq:planes_good}\\
    &\SMbR:=\{\MbH(K,\gamma,P_c)=\miM\},\nonumber
\end{align}
and we repeat the procedure for the standard approach, using $\suI$, $\suQ$, $\suM$,
so that the extracted surfaces read
\begin{align}
    &\SIW:=\{\ISH(K,\gamma,P_c)=\suI\},\nonumber\\
    &\SQW:=\{\QSH(K,\gamma,P_c,\Oms)=\suQ\},\label{eq:planes_bad}\\
    &\SMbW:=\{\MbH(K,\gamma,P_c)=\Mms\}.\nonumber
\end{align}

These surfaces are displayed in Figure~\ref{fig:grid_3d}. The comparison between columns (b) and (c) in Fig.~\ref{fig:grid_3d} shows that the use of the extended set of universal relations including Love-$\deltaM$
provides surfaces which, although not intersecting at $\pms$ (they may not even intersect at one point), pass much closer to the actual stellar model (shown in panel (a)) than those obtained
from the standard approach.

In order to obtain a quantitative estimation
of how much closer to $\pms$
are the polytropic EoS parameters inferred
by the extended approach as compared to the standard one,
we proceed as follows.
We define a dimensionless
distance between two points $p_1=(x_1,y_1,z_1)$ and $p_2=(x_2,y_2,z_2)$ in the space $\{K,\gamma,P_c\}$ as the Cartesian module of the vector
\begin{align}
    \vec{d}(p_1,p_2):=\left(\frac{ x_2-x_1}{\Kms},\frac{ y_2-y_1}{\gammams},\frac{z_2-z_1}{\Pcms}\right),\label{eq:3D_error}
\end{align}
so that we can define the shortest distances $\radio_X(p)$
from any point $p=(x,y,z)$ to the surface $\mathcal{S}_X$, 
with $X$ denoting any of the relevant quantities.

In particular, for the extended approach we define 
$\radiolms(p)$, $\radioMms(p)$, $\radiomiI(p)$, $\radiomiQ(p)$, $\radiomiM(p)$
as the shortest distance from $p$ to the surfaces $\Slove$, $\SMS$, $\SIR$, $\SQR$, $\SMbR$, respectively. 
One could consider different characterizations of the inferred solutions, considering for instance the point $p$ that minimizes the mean value of all the distances. However, to have an order-of-magnitude estimate we simply characterize the solution 
of the extended approach $\mipunto$ as the point that minimizes the distance to the furthest plane, i.e.~the value of $p$ for which
\begin{align}
    &\miradio(p):=\label{eq:3D_radius}\\
    &\mathrm{max}\{\radiolms(p),\radioMms(p),\radiomiI(p),\radiomiQ(p),\radiomiM(p)\},\nonumber
\end{align}
is minimum. In a sufficiently small neighbourhood of $\pms$ this will be well defined and will
provide a unique solution.
We can thus define the relative errors, corresponding
to the three EoS quantities, as the
components of the vector $\mierrorvector$ defined by
\begin{align}
    \mierror^{i}=\lvert d^{\,i}(\mipunto,\pms)\rvert.\label{eq:errores3d}
\end{align}

For the standard approach one only needs to change $\ext\rightarrow\std$ in the expressions \eqref{eq:3D_error}-\eqref{eq:3D_radius}. The relative errors $\mierrorK$, $\mierrorgamma$, $\mierrorPc$ and $\suerrorK$, $\suerrorgamma$, $\suerrorPc$ from the results of Fig.~\ref{fig:grid_3d}
are summarized in Table \ref{tab:relative_errors_3d}. As happens for the two-dimensional setup, the accuracy of the extended approach in the general setup is also significantly better. More precisely, the errors using the standard approach are $98-110$ times larger than those attained with the extended approach.

\begin{table}[t]
\begin{tabular}{cccc}
\hline
Approach & $\errors_K$       & $\errors_\gamma$  & $\errors_{P_c}$\\
         & $(\times10^{-4})$ & $(\times10^{-4})$ & $(\times10^{-4})$\\
\hline
Extended & 2 & 2 & 2 \\
Standard & 223 & 189 & 203 \\
\hline
\end{tabular}
\caption{Relative errors $\mierrorK$, $\mierrorgamma$, $\mierrorPc$ as defined in \eqref{eq:errores3d}, and the corresponding $\suerrorK$, $\suerrorgamma$, $\suerrorPc$.}
\label{tab:relative_errors_3d}
\end{table}

\subsection{MIT bag EoS: Inferring $B$ and $P_c$}\label{sec:mit_bag}

\begin{figure*}[htb]
\centering
    \includegraphics[width=\linewidth]{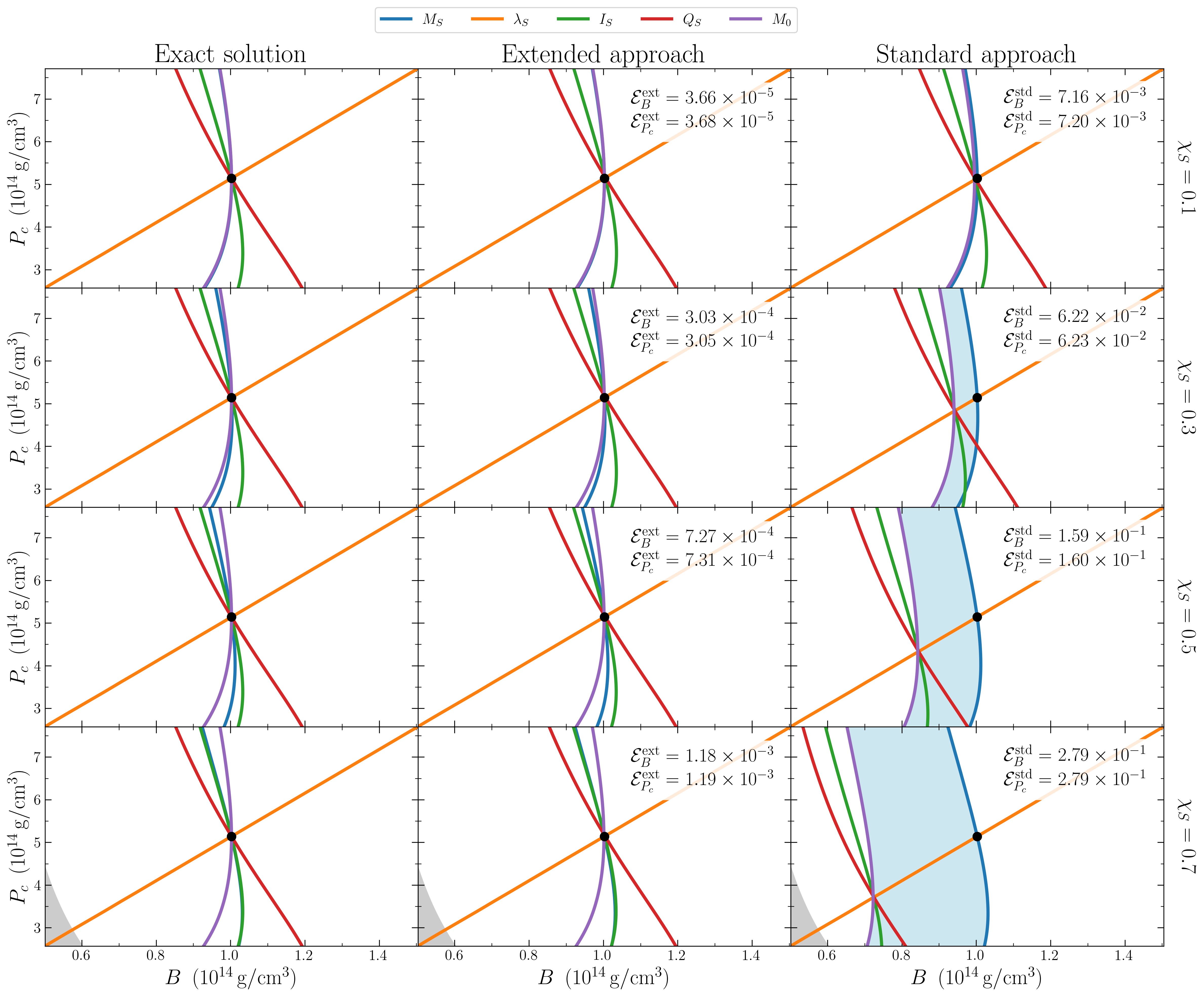}
\caption{As Fig.~\ref{fig:poly_K_fixed} but for the MIT bag EoS.
The stellar model has EoS parameters $\Bms=1.003\times10^{14}\,\mathrm{g}/\mathrm{cm}^3$ and $\Pcms=5.14\times10^{14}\,\mathrm{g}/\mathrm{cm}^3$. The two-dimensional grid spans the intervals $B\in[0.501,1.504]\times10^{14}\,\mathrm{g}/\mathrm{cm}^3$ and $P_c\in[2.57,7.71]\times10^{14}\,\mathrm{g}/\mathrm{cm}^3$.}
\label{fig:mitbag}
\end{figure*}

As our final model problem we now repeat the exercise in Sec.~\ref{sec:given_K} for the MIT bag stellar model. The procedure follows in a similar way, where the role of the pair $(\gamma, P_c)$ is now played by the pair $(B, P_c)$, that defines the grid. In particular, Eqs.~\eqref{eq:curves_exact}-\eqref{eq:error_Pc_W} still apply after a change $\gamma\rightarrow B$. The inference of $\Mb$, $\IS$ and $\QS$ in this case (as well as the actual $\Mms$ and $\lms$) are shown in Fig.~\ref{fig:mitbag} using both approaches. Again, we see that a more precise inference of the EoS parameters is obtained in the extended approach than in the standard one. The discrepancies in the latter for $B$ are approximately $196-236$ times larger than those in the extended version, and for $P_c$ they are about $196-235$ times larger. We also note that  for the MIT bag EoS the relative errors present significantly higher ratios than for the polytropic EoS. 

\subsection{Additional remarks}\label{sec:additional}

On an additional note, 
let us stress that for a given set of quantities $(\lms,\Mms,\Oms)$ the extended approach provides three additional quantities, namely $\miM$, $\miI$ and $\miQ$, whereas the standard procedure provides only two, $\suI$, $\suQ$ (which are moreover affected by the choice $\suM=\Mms$). As a result, the extended procedure may be better suited to determine the parameters describing more general EoS. 
As an example, let us assume we want to infer the EoS of a stellar configuration composed of two piecewise polytropic regions. Such type of constructions were proposed by~\cite{Read:2008iy} to parameterize a large family of cold EoS and is an approach routinely used in simulations of BNS mergers. Each region, labelled by $i=1,2$, is described by three free parameters, namely $K_{(i)}$, $\gamma_{(i)}$ and $P_{0(i)}$. In the central region, $P_{0(1)}$ corresponds to the central pressure, while $P_{0(2)}$ in the outer region is determined from the matching conditions, that demand the continuity of the pressure at the matching surface. This reduces the number of free parameters to five, which coincides with the number of quantities provided by the extended approach. Assuming there is no degeneracy on the problem, the extended approach could then be used to infer the five parameters of the EoS, whereas the standard approach would fail in the full determination of that relevant kind of models.

\section{Conclusions and discussion}\label{sec:conclusions}

The $I$-Love-$Q$ universal relations of rotating compact stars are relevant for observational astrophysics since they allow the inference of any two quantities within the triad out of the third one alone. In this paper we have discussed an extended version of the $I$-Love-$Q$ universal relations including a fourth parameter, $\delta M$, proportional to the difference between the mass of the rotating star $M_S$ and the mass of the (non-rotating) background configuration arising in perturbative studies, $M_0$.  In particular, $M_0$ plays a key role in the actual application of the relations, as it is used to normalize the star's moment of inertia and  quadrupole moment. The fact that $\Mb$ is a quantity that cannot be extracted from observations hinders the application of the relations, which is usually overcome by replacing it with the actual mass, $\MS$.
For slowly rotating compact stars, using $\MS$ instead of $\Mb$ only introduces small deviations in the predictions, since $\Mb\approx\MS$~\cite{Yagi:2013awa}.
For rapidly rotating stars, however, the mass $\MS$ differs significantly from its static and spherically symmetric counterpart, and non-negligible deviations from universality appear in the $I$-Love-$Q$ relations.

We have quantified the inaccuracies introduced by applying the universal relations under the assumption $M_0=M_S$. To this aim we have made use of our  extended set of universal relations $I$-Love-$Q$-$\delta M$, originally introduced in~\cite{reina-sanchis-vera-font2017}, as they allow to unambiguously extract $M_0$ with no approximation. Our tests have been performed using two different stellar models, a relativistic polytrope and the MIT bag EoS. In both cases our results highlight the importance of incorporating the second-order contribution of the mass $\deltaM$ into the  relations. The errors found on $\IS$, $\QS$ and $\Mb$ when the approximation $\Mb=\MS$ is made are consistently larger than when it is not. Moreover, those errors significantly increase with the rotation of the star, as expected.
We have also analyzed the inference of EoS parameters using the universal relations finding that the inference is greatly enhanced when using the extended set of relations.
It is worth noting that there exist other approaches that have considered $\chiS$-dependent ``universal" relations, which would apply to rapidly rotating stars as well (see~\cite{Doneva:2013rha,Pappas:2013naa,Chakrabarti:2013tca} for the $\chiS$-dependent $I$-$Q$ relations). However, the extended $I$-Love-$Q$-$\deltaM$ universal relations discussed in this work sidestep the use of such $\chiS$-dependent relations by enabling the inference of $\Mb$ in the first place.

We end this paper with some final remarks on the potential importance our extended set of universal relations might have if the sample of BNS systems with rapidly rotating components were to increase.
The approximation $\Mb=\MS$ was justified in~\cite{Yagi:2013awa} on the basis that neutron stars in binary systems are expected to rotate with dimensionless spin parameters $\chiS\lesssim 0.1$. Those values sharply contrast with the ones attained by millisecond pulsars, with the fastest spinning such pulsar known to date, PSR J1748-2446ad~\cite{Hessels:2006}, showing a spin period of 1.40 ms (i.e.~$\chiS\sim  0.4$). Moreover, the theoretical breakup spin of a uniformly rotating neutron star for realistic EoS is significantly larger, about $\chiS\sim 0.7$~\cite{Lo:2011}. Not surprisingly, recent works have argued whether the almost complete absence of millisecond pulsars in BNS systems could be affected by an observational bias~\cite{Papenfort:2021,Dudi:2022,Papenfort:2022,Rosswog:2023}.

Currently, about 23 BNS systems have been observed in our Galaxy, with about half of which expected to merge within a Hubble time~\cite{Tauris:2023}.  The pulsars observed in the fastest-spinning BNS systems of the sample, PSR J0737-3039A~\cite{Burgay:2003} and PSR J1946+2052~\cite{Stovall:2018}, both of which will merge within a Hubble time, will have dimensionless spins $\chiS\lesssim 0.05$ at merger. On the other hand, BNS systems in dense stellar environments like globular clusters may include a fully recycled millisecond pulsar because of the chance of close encounters among stars. Out of the 23 Galactic BNS systems three have been observed in globular clusters, PSR J0514-4002A~\cite{Ridolfi:2019}, PSR J1807-2459B~\cite{Lynch:2012}, and B2127+11C~\cite{Wolszczan:1989}. The first two host, indeed, a millisecond pulsar and the third one hosts a mildly recycled pulsar spinning at 30 ms. The latter is the only one of these three systems that will merge within a Hubble time. Recently~\cite{Rosswog:2023} have argued that about 5\% of all BNS mergers may actually contain one millisecond component. Moreover, triple systems may also contribute to the formation rate of rapidly spinning neutron stars in BNS systems~\cite{Hamers:2019}. It is worth pointing out that the ratio of BNS systems with one rapidly-rotating neutron star is expected to increase in the coming decade once the Square Kilometer Array (SKA) becomes operational. SKA anticipates about an order of magnitude growth in the number of Galactic BNS systems~\cite{Keane:2015}.

GW observations are also expected to shed light on the range of possible spins in BNS systems, particularly when the number of detections grows. Using the catalog of GW signals from compact binary mergers during the first three LIGO-Virgo observing runs, a rate of BNS mergers between 10 and 1700 Gpc$^{-3}$ yr$^{-1}$ has been inferred while the corresponding rate for neutron star-black hole mergers is between 7.8 and 140 Gpc$^{-3}$ yr$^{-1}$~\cite{GWTC3}. The LVK Collaboration also envisaged a detection count of $10^{+52}_{-10}$ for BNS mergers and of $1^{+91}_{-1}$ for neutron star–black hole mergers in a one-calendar-year observing run of the four-detector network during O4~\cite{LVK-observing}. For third-generation detectors, these figures will increase significantly. The expected BNS merger detection rates for the Einstein Telescope are $\sim  7\times 10^4$ binaries in one year~\cite{Belgacem:2019}, based on the ET-D design sensitivity~\cite{Hild:2011}. Correspondingly,  Cosmic Explorer is expected to detect $\sim 10^5$ BNS mergers per year out to a redshift of $z=4$~\cite{Evans:2021}.

With so many anticipated detections, it is essential to develop approaches able to describe generic BNS systems to aid on the detection and on the parameter inference and data analysis, i.e.~including systems in which the individual neutron stars are millisecond pulsars. Over the last few years different groups have begun to study BNS configurations in which the individual stars are spinning, including also highly spinning models (see~\cite{Dudi:2022} and references therein). In particular, the numerical-relativity simulations of~\cite{Dudi:2022} show that for the highest spinning configurations analyzed, corresponding to $\chiS = 0.58$ and $\chiS = -0.28$, existing waveform models employed for the interpretation of signals GW170817 and GW190425 fail to provide an accurate description during the late inspiral. Moreover, given the degeneracy on the inspiral waveform between effects induced by the tidal deformations and by the spins, the mass ratio between the constituents of GW170817 and GW190425 is not well constrained, depending on whether low-spin or high-spin priors are applied~\cite{GW170817-properties,GW190425}, with the latter yielding a larger uncertainty. Signal GW190814~\cite{GW190814} also provides further support for the possible existence of rapidly-rotating neutron stars in compact binary systems. The possibility that the secondary is a mass-gap, fast-rotating neutron star has been studied by~\cite{Zhang:2020,Most:2020,Biswas:2021}, with an estimation of the  dimensionless spin of $\chiS\sim 0.47$~\cite{Biswas:2021} and $\chiS\sim 0.49$~\cite{Most:2020}. This is about 65\% of the maximum spin ($\chiS\sim 0.7$) uniformly rotating isolated neutron stars can reach~\cite{Lo:2011}. 
In this context, it is also worth pointing out the work by~\cite{Kuan:2022} where it was noted that the dephasing of the inspiral GW signal by the excitation of the $f$-mode in coalescing binaries containing one neutron star exceeds the uncertainty in the phase of the waveform when the neutron star rotates above 800 Hz, which renders necessary the incorporation of high spins in the waveform templates. In summarize, the (few) LVK observations and analysis of BNS mergers to date have left the pre-merger spins of the neutron stars essentially unconstrained.

Unmistakably, the previous discussion motivates the development of approaches to describe
generic BNS systems,  including, in particular, systems in which the individual neutron stars are rapidly-rotating. The extended set of universal relations discussed in this work could become useful for the determination of neutron star properties in such circumstances.


\begin{acknowledgments}
We are thankful to Daniela Doneva for interesting comments. Work supported by the Spanish Agencia Estatal de Investigaci\'on (grants PID2021-125485NB-C21, PID2021-123226NB-I00 funded by MCIN/AEI/10.13039/501100011033 and ERDF A way of making Europe), by the Generalitat Valenciana (grant CIPROM/2022/49), by the Basque Government (IT1628-22), and by the European Horizon Europe staff exchange (SE) programme HORIZON-MSCA-2021-SE-01 (NewFunFiCo-101086251). EA is supported by the Basque Government Grant No.~PRE\_2023\_2\_0199. NSG acknowledges support from the Spanish Ministry of Science and Innovation via the Ram\'on y Cajal programme (grant RYC2022-037424-I), funded by MCIN/AEI/10.13039/501100011033 and by ``ESF Investing in your future”.  
\end{acknowledgments}

\appendix

\section{$\deltaM$-$I$ and $\deltaM$-$Q$ relations}\label{app:fittings}

Although the prescription to compute $I$, $Q$ and $\deltaM$ from $\lS$ only requires the set of universal relations summarized in Table~\ref{tab:values_fitting}, we present in this appendix the fitting parameters of the remaining $\deltaM$-$I$ and $\deltaM$-$Q$ relations,
which, to our knowledge, have not been yet reported.

The fitting relations have been computed for various stellar models.
On the one hand, we have considered two-fluid superfluid neutron stars with different EoSs: 
GM1 \cite{Glendenning:1991} and NL3 \cite{Fattoyev:2010} are different parametrizations of the relativistic mean field model EoS from \cite{Kheto2015}, Poly corresponds to the EoS from \cite{comer1999} where each fluid behaves as a relativistic polytrope, and TOY is the toy model presented in \cite{Aranguren:2022} that is based on the polytropic EoS.
Next, we have also considered the standard MIT bag model developed in \cite{chodos_1:1974,chodos_2:1974,degrand:1975} with 4 parametrizations of the EoS: Colpi $\&$ Miller \cite{Colpi:1992}, and MITBAG1-3 which correspond to the SQM1-3 models in \cite{Lattimer_2001} computed using the linear EoS proposed in \cite{Zdunik:2000}.
Finally, we have also considered some tabulated EoS, namely EoS A from \cite{pandharipande:1971A} and EoS FPS from \cite{Lorenz_EOS}.

\begin{figure*}[ht]
    \centering
    \includegraphics[width=\textwidth]{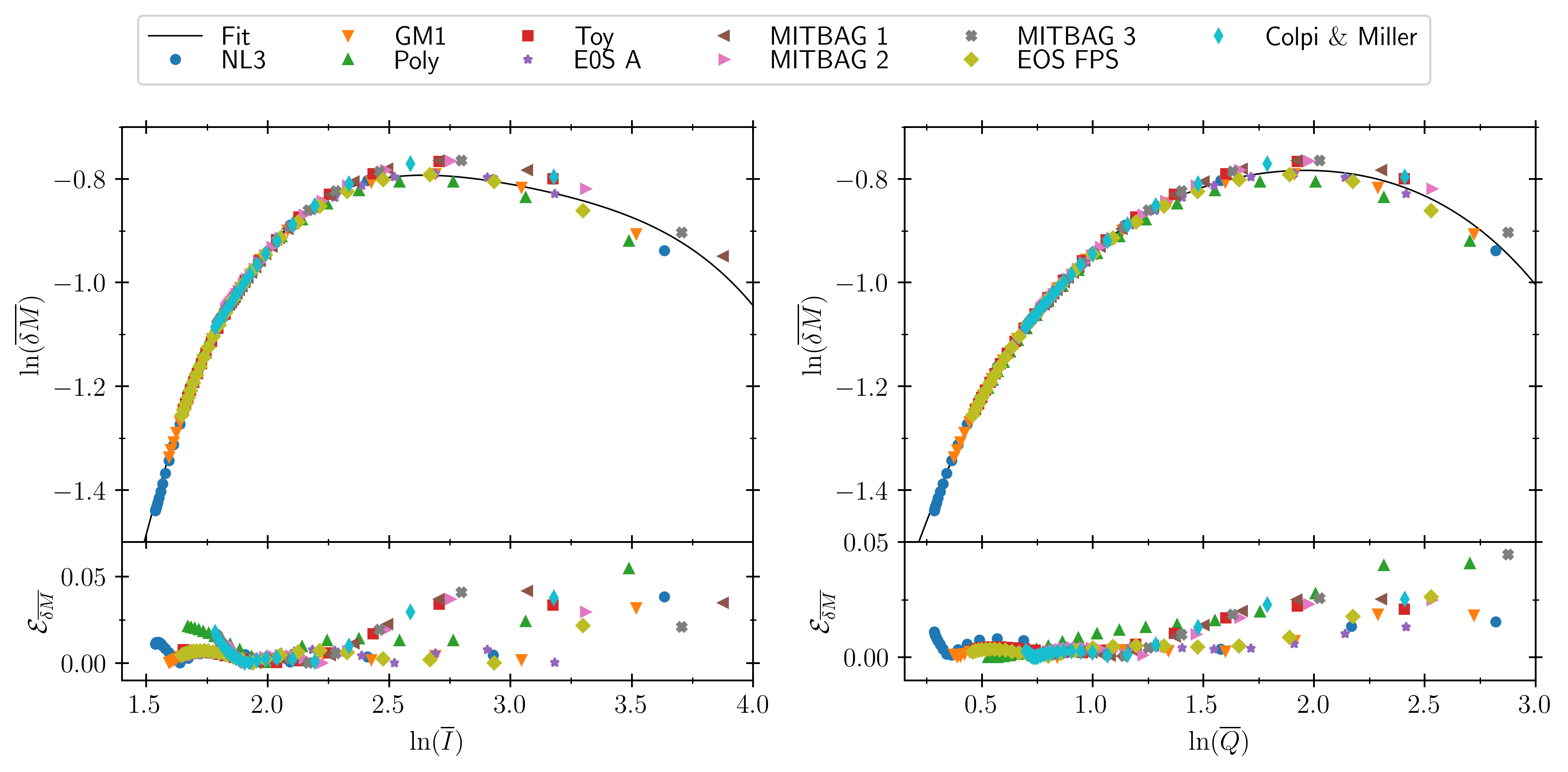}
    \caption{Least squares fits for the $\deltaM$-$I$ and $\deltaM$-$Q$ universal relations for various single fluid and two fluid EoS, truncated at order four. The lower panel in each plot shows the relative errors $\mathcal{E}_X=|(\ln X-\ln X^{\mathrm{fit}})/\ln X^{\mathrm{fit}}|$.}
    \label{fig:dMIdMQ}
\end{figure*}

The numerical results for the $\deltaM$-$I$ and $\deltaM$-$Q$ universal relations are displayed in Fig.~\ref{fig:dMIdMQ}. The  corresponding set of coefficients  characterizing the fitting curve from equation \eqref{eq:fittingcurve} (displayed as a solid line in Fig.~\ref{fig:dMIdMQ}) are summarized in Table \ref{tab:values_fitting2}.

\begin{table}[t]
\setlength{\tabcolsep}{2.1pt}
\renewcommand{\arraystretch}{1.5}
    \centering
    \begin{tabular*}{\linewidth}{@{\extracolsep{\fill}} c c c c c c c }
    \hline\hline
         $y_i$ & $x_i$ & $a_i$ & $b_i$ & $c_i$ & $d_i$ & $e_i$ \\
    \hline
         $\overline{\deltaM}$ & $\overline{I}$ & $-10.51$  & $12.46$ & $- 5.985$ & $1.283$ & $ -0.1044$ \\
         $\overline{\deltaM}$ & $\overline{Q}$ & $-1.784$  & $1.502$ & $- 0.888$ & $0.2645$ & $- 0.0355$ \\
    \hline\hline
    \end{tabular*}
    \caption{Best fitting parameters for the curves in Fig.~\ref{fig:dMIdMQ}.}
    \label{tab:values_fitting2}
\end{table}
